\documentclass[aps,prl,
superscriptaddress,
twocolumn,
reprint
]{revtex4-1}
\usepackage{color}
\usepackage{graphicx}
\usepackage{amsmath}
\usepackage{amsfonts}
\usepackage{amssymb}
\usepackage{color}
\usepackage[utf8]{inputenc}
\usepackage[T1]{fontenc}
\usepackage[italian,english]{babel}
\usepackage{hyperref}

\newcommand{\beq}{\begin{equation}} 
\newcommand{\eeq}{\end{equation}}
\newcommand{\bfig}{\begin{figure}}
\newcommand{\efig}{\end{figure}}
\newcommand{\igraph}{\includegraphics}

\begin{document}

\title{Tuning of magnetic activity in spin-filter Josephson junctions towards spin-triplet transport}
\author{R. Caruso}
\affiliation{Dipartimento di Fisica E. Pancini$,$ Università degli Studi di Napoli Federico II$,$ Monte S. Angelo$,$ via Cintia$,$ I-80126 Napoli$,$ Italy}
\affiliation{CNR-SPIN$,$ UOS Napoli$,$ Monte S. Angelo$,$ via Cintia$,$ I-80126 Napoli$,$ Italy}
\author{D. Massarotti}
\affiliation{Dipartimento di Ingegneria Elettrica e delle Tecnologie dell'Informazione$,$ Università degli Studi di Napoli Federico II$,$ via Claudio$,$ I-80125$,$ Italy}
\affiliation{CNR-SPIN$,$ UOS Napoli$,$ Monte S. Angelo$,$ via Cintia$,$ I-80126 Napoli$,$ Italy}
\author{G. Campagnano}
\affiliation{Dipartimento di Fisica E. Pancini$,$ Università degli Studi di Napoli Federico II$,$ Monte S. Angelo$,$ via Cintia$,$ I-80126 Napoli$,$ Italy}
\affiliation{CNR-SPIN$,$ UOS Napoli$,$ Monte S. Angelo$,$ via Cintia$,$ I-80126 Napoli$,$ Italy}
\author{A. Pal}
\affiliation{Department of Materials Science and Metallurgy$,$  University of Cambridge$,$ 27 Charles Babbage Road$,$ Cambridge CB3 0FS$,$ UK}
\author{H. G. Ahmad}
\affiliation{Dipartimento di Fisica E. Pancini$,$ Università degli Studi di Napoli Federico II$,$ Monte S. Angelo$,$ via Cintia$,$ I-80126 Napoli$,$ Italy}
\affiliation{CNR-SPIN$,$ UOS Napoli$,$ Monte S. Angelo$,$ via Cintia$,$ I-80126 Napoli$,$ Italy}
\author{P. Lucignano}
\affiliation{Dipartimento di Fisica E. Pancini$,$ Università degli Studi di Napoli Federico II$,$ Monte S. Angelo$,$ via Cintia$,$ I-80126 Napoli$,$ Italy}
\affiliation{CNR-SPIN$,$ UOS Napoli$,$ Monte S. Angelo$,$ via Cintia$,$ I-80126 Napoli$,$ Italy}
\author{M. Eschrig}
\affiliation{Department of Physics$,$ Royal Holloway$,$ University of London$,$ Egham$,$ Surrey TW20 0EX$,$ UK}
\author{M. G. Blamire}
\affiliation{Department of Materials Science and Metallurgy$,$  University of Cambridge$,$ 27 Charles Babbage Road$,$ Cambridge CB3 0FS$,$ UK}
\author{F. Tafuri}
\affiliation{Dipartimento di Fisica E. Pancini$,$ Università degli Studi di Napoli Federico II$,$ Monte S. Angelo$,$ via Cintia$,$ I-80126 Napoli$,$ Italy}
\affiliation{CNR-SPIN$,$ UOS Napoli$,$ Monte S. Angelo$,$ via Cintia$,$ I-80126 Napoli$,$ Italy}

\begin{abstract}
The study of superconductor/ferromagnet interfaces has generated a great interest in the last decades, leading to the observation of equal spin spin triplet supercurrent and $0-\pi$ transitions in Josephson junctions where two superconductors are separated by an itinerant ferromagnet.
Recently, spin-filter Josephson junctions with ferromagnetic barriers have shown unique transport properties, when compared to standard metallic ferromagnetic junctions, due to the intrinsically non-dissipative nature of the tunneling process.
Here we present the first extensive characterization of spin polarized Josephson junctions down to 0.3 K, and the first evidence of an incomplete $0-\pi$ transition in highly spin polarized tunnel ferromagnetic junctions.
Experimental data are consistent with a progressive enhancement of the magnetic activity with the increase of the barrier thickness, as neatly captured by the simplest theoretical approach including a non uniform exchange field.
For very long junctions, unconventional magnetic activity of the barrier points to the presence of spin-triplet correlations.
\end{abstract}

\maketitle

The interaction of superconductors with materials other than simple insulators or metals has made accessible a series of conceptually new challenges. 
Of particular interest to this work, Josephson junctions (JJs) with ferromagnetic materials separating two superconductors have been extensively characterized over the last decade. The simultaneous presence of the macroscopic phase coherence of superconductors and the exchange interaction of ferromagnetic materials is indeed of great value in the study of fundamental questions on possible pairing states in superconductors\cite{Bergeret:2005-Review, Buzdin:2005}, demonstrating the presence of spin-polarized triplet supercurrents \cite{triplet1, triplet2, triplet3, triplet4, triplet5, triplet6, Birge:2016}, and for potential applications in a wide range of cutting edge areas, such as spintronics \cite{Eschrig:2015-Spintronics, Linder:2015}, memory applications for high performance computing \cite{mem1, memBaek, memorieJAP, memorieTAS, memGoldobin, memBirge, memBirge2018} and circuit components such as $\pi$ shifters and phase qubits \cite{qbit:ustinov, qbit:ustinov2, qbit:yamashita, sfq:ryazanov2010, Bolginov:2018}. A playground where different forms of order can cooperate and interfere is of considerable value for inspiring other fields of physics \cite{Bergeret:2005-Review, Buzdin:2005}.

The existing literature focuses mostly on metallic superconductor/ferromagnet/superconductor (SFS) junctions, where the evidence of long-range spin triplet correlations is well established \cite{triplet1, triplet2, triplet3, triplet4, triplet5, triplet6}: in the presence of equal-spin Cooper pairs, the magnitude of the critical current $I_C$ decays much more slowly with magnetic barrier thickness than expected for standard singlet supercurrents \cite{triplet2, triplet3}. In fact, spin-polarized Cooper pairs can survive at much longer length scales when compared to opposite spin Cooper pairs, and are practically immune to depairing induced by the presence of an exchange field \cite{Bergeret:2005-Review, Buzdin:2005}. Such junctions, together with superconducting spin valve devices, are likely to be the building blocks for future spintronic devices \cite{Linder:2015}. While metallic SFS junctions have been extensively characterized, the physics of ferromagnetic junctions with insulating barriers, like the ones in this work, is still relatively unexplored, despite the unique key feature of falling in the underdamped regime.

Recent results on GdN/Nb/GdN \cite{Zhu:2017} have revealed the presence of a novel exchange interaction between ferromagnetic insulator GdN layers, mediated by the Nb interlayer, thus promoting possible control of the magnetic state in spin valve structures by superconductivity. Our work focuses on NbN/GdN/NbN spin filter (SI$_{\rm F}$S) Josephson junctions: spin filter JJs have emerged as an extremely promising solution in the field in the last few years \cite{Senapati:2011}, since the insulating nature of the ferromagnetic barrier promotes higher values of the $I_C R_N$ product, up to 1 mV ($R_N$ being the normal state resistance), when compared to conventional metallic SFS JJs, and provides the first evidence of macroscopic quantum phenomena in SFS JJs \cite{davideNATCOM}. 

In this work, we present measurements of the properties of spin filter junctions as a function of temperature $T$ for different values of the thickness $d$ of the GdN layer in a wider range of $T$ and $d$ with respect to \cite{Pal:2014}. 
These junctions are known to exhibit unconventional $I_C(H)$ behavior at 4.2K as extensively reported in \cite{Pal:2014} (see Supplementary Material for further details). When investigated at lower temperatures and higher spin filter efficiencies, they show properties which are only consistent with a unique magnetic activity of the barrier. In metallic ferromagnetic junctions, the presence of a magnetic activity modeled by a non-uniform exchange field and a spin filtering effect has been related to the presence of triplet correlations contributing to the total supercurrent \cite{Bergeret:2005-Review, triplet2}. 
Here we show that for long SI$_{\rm F}$S junctions non-uniform exchange fields as well as spin-triplet correlations play an important role for explaining the $I_C(T)$ experimental data.

We observe a net deviation from the expected Ambegaokar-Baratoff behavior\cite{AB:1963}, which becomes dramatically evident for samples with barrier thicknesses above 2.5 nm, where a critical current plateau appears at intermediate temperatures, indicating an incomplete $0-\pi$ transition\cite{piJJ, piJJ:aprili}. 
The complete transition has been widely observed in metallic SFS junctions, and it is attributed to the oscillating behavior of the superconducting order parameter inside an itinerant ferromagnet, due to the presence of the homogeneous exchange field of the ferromagnet.
This transition has been theoretically predicted also for SI$_{\rm F}$S junctions by Kawabata et al. \cite{Kawabata:2010}, but to the best of our knowledge it has never been observed experimentally. Our experimental observation of an incomplete $0-\pi$ transition cannot be explained using the standard proximity effect with a conventional s-wave order parameter. Only the assumption of interfacial inhomogeneities, together with the presence of a high spin-filter efficiency and a large spin mixing angle allows to obtain a good agreement between experimental data and theoretical curves.

The junctions have been fabricated in the same fabrication run, varying GdN thickness by changing the deposition rates \cite{Pal:2014}, in order to ensure the same deposition conditions for all samples. All the measurements have been performed using an evaporation cryostat with a base temperature of 300 mK with customized low noise filters anchored at different temperature stages\cite{luigiAPL2011, davidePRB2015, danielaLAOSTO2017}. More details can be found in Supplementary Material.

Spin-filter properties of these junctions have been extensively discussed in \cite{Senapati:2011}. 
The values of the spin-filter efficiency $P$ for the measured junctions are shown in Fig.\ref{fig:pars} (a): $P$ increases as the thickness increases, and saturates for barrier thicknesses above 3 nm, corresponding to a spin filter efficiency larger than 95\%.
\begin{table}[t]
\centering
\begin{tabular}{|p{0.180\columnwidth}|p{0.180\columnwidth}|p{0.180\columnwidth}|p{0.180\columnwidth}|p{0.180\columnwidth}|}
\hline d (nm) &  $I_C$   &  $t_{\uparrow}$ & $t_{\downarrow}$ & $\Theta$ \\
\hline 1.5 & 680 $\mu$A  & 0.107 & 0.198 & 0 \\
\hline 1.75 & 220 $\mu$A & 4.95$\times 10^{-2}$ & 0.215 & 0.470 \\
\hline 2 &  250 $\mu$A  & 3.94$\times 10^{-2}$ & 0.206 & 0.400  \\
\hline 2.5 & 40 $\mu$A & 8.32$\times 10^{-3}$ & 0.132 & 0.935 \\ \hline
\end{tabular}
\begin{tabular}{|p{0.055\textwidth}|p{0.055\textwidth}|p{0.090\textwidth}|p{0.090\textwidth}|p{0.05\textwidth}|p{0.04\textwidth}|p{0.045\textwidth}|}
\hline d (nm) & $I_C$ & $t_{\uparrow}$ & $t_{\downarrow}$ & $\Theta_1$ & $\Theta_2$ & $g$ \\
\hline 3 & 5.2 $\mu$A & 2.33$\times 10^{-3}$ & 0.157 & 1.992 & 3.10 & 0.400 \\
\hline 3.5 & 590 nA & 5.40$\times 10^{-4}$ & 5.90$\times 10^{-2}$ & 2.256 & 3.14 & 0.292 \\
\hline 4 & 30 nA & 2.82$\times 10^{-4}$ & 3.90$\times 10^{-2}$ & 2.085 & 2.95 & 0.400 \\ \hline
\end{tabular}
\caption{Junction parameters at 0.3 K and fitting parameters for the samples analyzed in this work. Barrier thickness $d$, critical current $I_C$, transparencies $t_\uparrow$ and $t_\downarrow$, spin mixing angles $\Theta_1$ and $\Theta_2$, and the ratio between the two transport channels $g$.}
\label{tab:tab1}
\end{table}

\bfig
\centering
\igraph[width=\columnwidth]{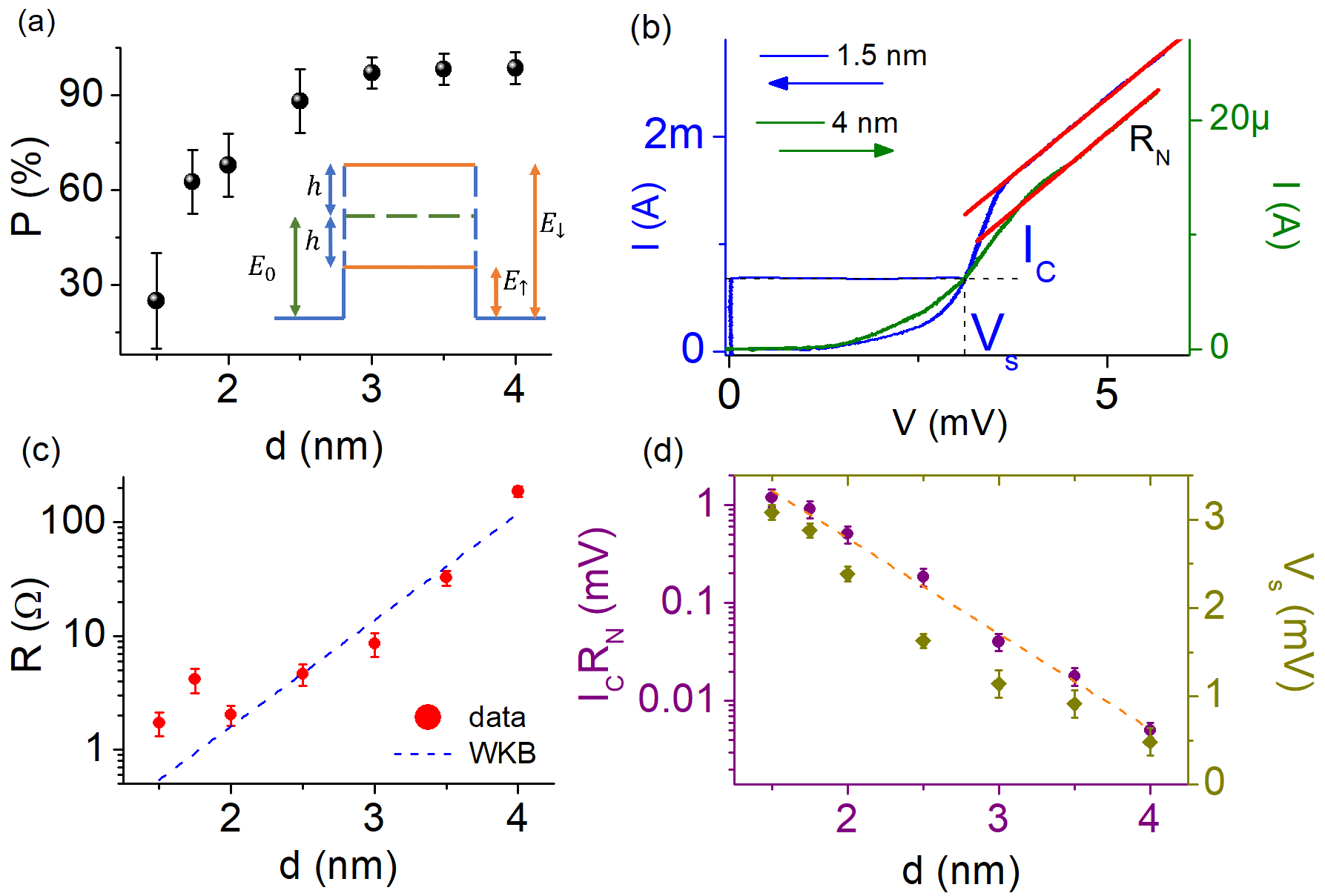}
\caption{(a) Spin filter efficiency at 15K as a function of thickness. Inset: sketch of the spin splitting of the barrier. (b) Typical I-V characteristic for NbN-GdN-NbN junctions at 0.3 K. Green: 4 nm thick barrier, blue: 1.5 nm thick barrier. The voltage is reported on the horizontal axis, while different scales for current are used on the vertical axes. Labels refer to the blue curve and indicate the characteristic parameters: 
the switching voltage $V_S$, the normal state resistance $R_N$ and the critical current $I_C$.
(c) $R_N(d)$ at 0.3 K, red dots are experimental data, blue dashed line is WKB fit. (d) $I_C R_N(d)$ (purple dots, bottom left axes) and $V_S(d)$ (dark yellow diamonds, top right axes) both measured at 0.3 K as a function of barrier thickness; orange dashed line is an exponential fit of $I_C R_N$ using Ref. \cite{birge:AF:PRB2009} with the GdN mean free path as fitting parameter.}
\label{fig:pars}
\efig
The study of the transport properties in spin filter junctions has been carried out theoretically in \cite{Bergeret:2012PRB}, in the following we focus our attention on the analysis of experimental current-voltage characteristics [$I(V)$] of these junctions.
In Fig. \ref{fig:pars} (b) we show typical $I(V)$ curves of the samples analyzed throughout this work. 
They refer to the extreme cases of $d = 1.5 $ nm (blue curve) and $d = 4 $ nm (green curve). The amplitude of the hysteresis is always larger than 90\% of the total current and the subgap leakage currents are larger for increasing $d$, consistently with the tunnel junction microscopic (TJM) model \cite{barone}.
$R_N$ increases as thickness increases (Fig. \ref{fig:pars} (c), red dots), in good agreement with the predictions within the Wentzel-Kramers-Brillouin (WKB) approximation \cite{bohm, messiah}, valid for standard tunnel junctions (blue dashed curve in Fig. \ref{fig:pars} (c)).
The $I_C R_N$ product and the switching voltage $V_{S}$ are also reported as a function of barrier thickness in Fig. \ref{fig:pars} (d). The exponential decay of the $I_C R_N$ product is consistent with the presence of different barrier heights seen by different spin channels, as in inset of Fig. \ref{fig:pars} (a) \cite{Senapati:2011}.
It has been shown \cite{Bergeret:2001, Kashuba:2007, birge:AF:PRB2009} that for metallic SFS junctions in the intermediate regime between the pure ballistic case and the diffusive limit, $I_C R_N$ decays exponentially with increasing thickness, with a decay constant equal to the electron mean free path in the ferromagnetic barrier. Using this model, we obtain a decay length in GdN $\xi \approx 0.4 $ nm, which is an unphysically low value, far lower than $\xi = 11 $ nm reported in literature for heavily doped, semiconducting GdN \cite{Ludbrook:2009}. This supports our assumption of a ballistic barrier, with an effectively insulating GdN, as discussed in \cite{Pal:2013}, where the transport cannot be described in terms of proximity effect but has to be modeled in terms of Cooper pair tunneling through the barrier.

In Fig. \ref{fig:pars} (d), $V_S$ decays linearly with increasing thickness.
Differently from standard tunnel junctions where $V_{S} \approx I_C R_N$, in spin filter junctions this correspondence does not hold anymore because of the strong $I_C$ suppression with increasing spin filter efficiency. The linear decrease of $V_{S}$ is a consequence of the subgap region of the $I(V)$ curves of spin filter junctions, which has a smooth dependence on voltage [see Fig. \ref{fig:pars} (b), (c) and (d)].
The junction parameters for all the samples analyzed in this work are collected in Table \ref{tab:tab1}.
Our measurements on $P$ and $I_C R_N$ as a function of thickness are in good agreement with previous literature on spin filter junctions \cite{Pal:2014}.

The $I_C (T)$ curves have a systematic dependence on $d$, which can be analyzed by distinguishing two different regimes. 
The first regime holds for thickness up to 2.5 nm, here $I_C$ presents  a progressive deviation from the conventional Ambegaokar-Baratoff behavior [Fig.\ref{fig:fitIcSum} (a)-(d)]. A monotonous decrease of $I_C$ occurs at higher temperatures, with a weak $T$ dependence at lower temperatures and a steeper decrease above 0.5 $T_C$. 
In the second regime, for barrier thicknesses larger than 3 nm, where the spin filtering properties are more relevant, $I_C$ presents an unconventional temperature dependence, with a clear non-monotonic behavior at large barrier thicknesses. The plateau, which extends from roughly 0.3 $T_C$ to 0.8 $T_C$ [see Fig.\ref{fig:fitIcSum} (e), (f)] in the junctions with 3 nm and 3.5 nm barrier thicknesses, evolves into a peak structure at about $T= 0.7 T_C$ for the junction with $d$ = 4 nm [see Fig.\ref{fig:fitIcSum} (g)]. 

This behavior does not have any analogy in literature and cannot be explained by any of the common theories\cite{AB:1963, LikharevRev:1979, Bergeret:2005-Review, Buzdin:2005}. We have developed a simple model to describe the junction behavior and in particular the $I_C(T)$  to a good approximation, and unambiguously correlate it to specific parameters describing the magnetic properties of the barrier, namely the presence of spin-filtering and spin-mixing. 

Given the insulating nature of the barrier \cite{Senapati:2011, Pal:2013}, we can assume that the samples are described within a ballistic transport theory. Typical values for $\xi_0$ in NbN at 4.2 K are between 3 nm and 5 nm \cite{xi1} so we can also assume that at least junctions with barrier thickness $d\leq 2.5 $ nm are in the short junction limit in the whole temperature range.
In this case the current is carried solely by the subgap Andreev levels that we calculate by neglecting the induced correlations in the GdN, as well as a possible renormalization of the NbN s-wave singlet order parameter due to the proximity effect. 
In the following, we explicitly exclude the contribution from equal spin Cooper pairs, because the large uncertainty of the magnetization profile in the ferromagnetic barrier does not allow its calculation. A rigorous microscopic model would also require to include a distribution of Andreev channels, due to the inhomogeneous properties of the magnetization as well as transmission characteristics for different interface regions. In this case, the subgap resonances due to Andreev states will be merged and smeared out into a continuum, as observed in conductance measurements (see Supplementary Material for details).
Hence, here we resort to a simplified description, including one or two spin dependent transport channels attempting to describe the relevant physics of the device, in particular the influence of its magnetic properties on the supercurrent.
The Josephson current can be expressed as:
\begin{equation}
I(\varphi)=-\frac{e}{\hbar}\sideset{}{}\sum_{\varepsilon<0}  \frac{\partial \epsilon_{n}(\varphi)}{\partial \varphi} \tanh{\left(\frac{\epsilon_{n}}{2k_B T}\right)}
\;\;\; , 
\label{eqn:cpr}
\end{equation}
where 
$T$ is the temperature and the sum is taken only on the negative (occupied) energy Andreev levels $\epsilon_n$, calculated following \cite{Beenakker:1991} and \cite{Tokuyasu:1998}. The details of the calculation can be found in the Supplementary Material.

The spin filtering is provided by allowing different transparencies $t_\uparrow=t \sin (\gamma)$ for spin up and $t_{\downarrow}=t \cos(\gamma)$ for  spin down electrons.
The angle $\gamma$ varies between 0 and $\pi/4$, and it is derived from the measured spin filter efficiency.
The effect of the exchange field is modeled as a spin-mixing angle $\Theta$, which accounts for the spin dependent scattering phases. These two angles, $\gamma$ and $\Theta$, are the key parameters of the model, as they allow to describe the effect of the ferromagnet on the transport properties of the junction \cite{Eschrig:2011}.
The sum in Eq.~\ref{eqn:cpr} is performed over four discrete subgap Andreev levels \cite{Eschrig:2000,Fogelstrom:2000,Barash:2002}, the first two are given by:
\begin{equation}
\varepsilon_{\pm}=|\Delta| \mbox{sgn}(\sin\frac{\Phi_{\pm}}{2})\cos\frac{\Phi_{\pm}}{2},
\end{equation} 
with 
\begin{equation}
\Phi_{\pm}(\varphi)=\Theta \pm \arccos[\sqrt{(1-t_\uparrow)(1-t_\downarrow)}-\sqrt{t_\uparrow t_\downarrow} \cos{\varphi}].
\end{equation}
The other two Andreev states are obtained by exchanging $\Theta \rightarrow -\Theta$.
In Fig. \ref{fig:fitIcSum} (a)-(d) we show the experimental data and the fitting curves obtained within this model. For $1.5 $ nm$ \leq d \leq 2.5$ nm, the experimental data (black dots) can be well fitted over the entire temperature range.  
Such curves deviate from the conventional Ambegaokar-Baratoff approach (blue dashed curve, plotted for comparison).  Our results show an unconventional $I_C(T)$ behavior even in this short limit. 
The single spin mixing angle for such junctions indicates a uniform, relatively small exchange field, while the transparencies $t_{\uparrow(\downarrow)}$ diminishes as the barrier thickness increases.
In Table \ref{tab:tab1} we report the fitting parameters for each experimental dataset.

The model used for short junctions fails to describe longer junctions with very high spin filter efficiency in the whole temperature range.
The change in slope of $I_C(T)$ curves for long junctions (from 3 nm to 4 nm) points to an incomplete $0-\pi$ transition, which could not be reproduced by the short junction approximation model described above, even including relaxation terms and the current contribution arising from the continuum part of the energy spectrum \cite{Brouwer:1997}. Green dash-dotted lines in Fig.\ref{fig:fitIcSum} (e)-(f) are theoretical curves showing  a complete $0-\pi$ transition obtained by assuming spin filter efficiency, transparencies and spin mixing angle expected for junctions with barrier thicknesses $3 $ nm $\leq d \leq 4 $ nm.

As a first approximation, the smoothing of the $0-\pi$ transition observed experimentally can be attributed to the magnetic structure of the barrier.
The GdN barrier has a fine internal domain structure due to the large area ($7\mu$m$ \times 7\mu$m) of the junctions\cite{Pal:2013, BlamireIcH}, as confirmed by the magnetization reversal behavior, calculated using the methods described in \cite{Jmagn}, in substantial agreement with the results shown in \cite{SenapatiPRB2011} for unpatterned GdN thick films (see Supplementary Material for further details). 
This structure influences the interface properties of the samples, giving rise to a non-uniform distribution of the ferromagnetic features of the barrier, that may be modeled as composed by different transport channels. 
We consider the simplest case of two channels each characterized by the same $t_\uparrow$, $t_\downarrow$ parameters, with $t_\uparrow\neq t_\downarrow$, and different spin mixing angles $\Theta_1$ and $\Theta_2$, with different weights reflecting the complex structure of the barrier. The critical current is obtained by maximizing the sum over all the transport channels. In this case one has:
\beq
I_C (T)= max_{\varphi}\left[ I(\varphi, \Theta_1)+g I(\varphi,\Theta_2)\right]
\label{eqn:ictsum}
\eeq
where $I$ is given by Eq.\ref{eqn:cpr} and $g=N_2/N_1$ is the relative weight of the two channels.
Each of the two channels can in principle undergo a $0-\pi$ transition, but only when combined together through Eq. \ref{eqn:ictsum} they give rise to the smoothed transition observed in the experiment.
In Fig.\ref{fig:fitIcSum} we show the fitting results obtained within this framework. The agreement between calculated $I_C(T)$ curves and measured points is significantly improved with respect to the single channel short junction model.
In Table \ref{tab:tab1} we report the fitting parameters obtained for long junctions.

We expect that increasing the number of transport channels, i.e. adding other terms to Eq. (\ref{eqn:ictsum}), can improve the agreement between experimental data and model, without adding any further contribution to the physical picture of the system. As in the case of a single channel, we find that the inclusion of other physical mechanisms to the model, such as broadening, relaxation and contributions from the continuum part of the energy spectrum, does not improve the agreement with experimental curves. In other words, exploiting this model to its maximum, our measurements can be explained only if we consider an increasing complexity of the magnetic activity of the barrier. Following the analogy with metallic, diffusive ferromagnetic systems, where the presence of spin-active interfaces implies the presence of equal spin triplet correlations, a promising route to model the transport properties in SI$_F$S junctions would be to further explore the role of spin-triplet correlations across the barrier, which is tightly connected with its complex magnetic structure. 

This phenomenology is complementary to what observed in junctions with itinerant ferromagnets as weak links, like in \cite{triplet2, triplet3}. In our case, the weak link is a tunnel barrier, and thus we do not expect to observe the slow decay of $I_CR_N$ typical of other SFS junctions in presence of triplet supercurrents, but rather an exponential decay with a decay length determined by the barrier height.

\bfig
\centering
\igraph[width=\columnwidth]{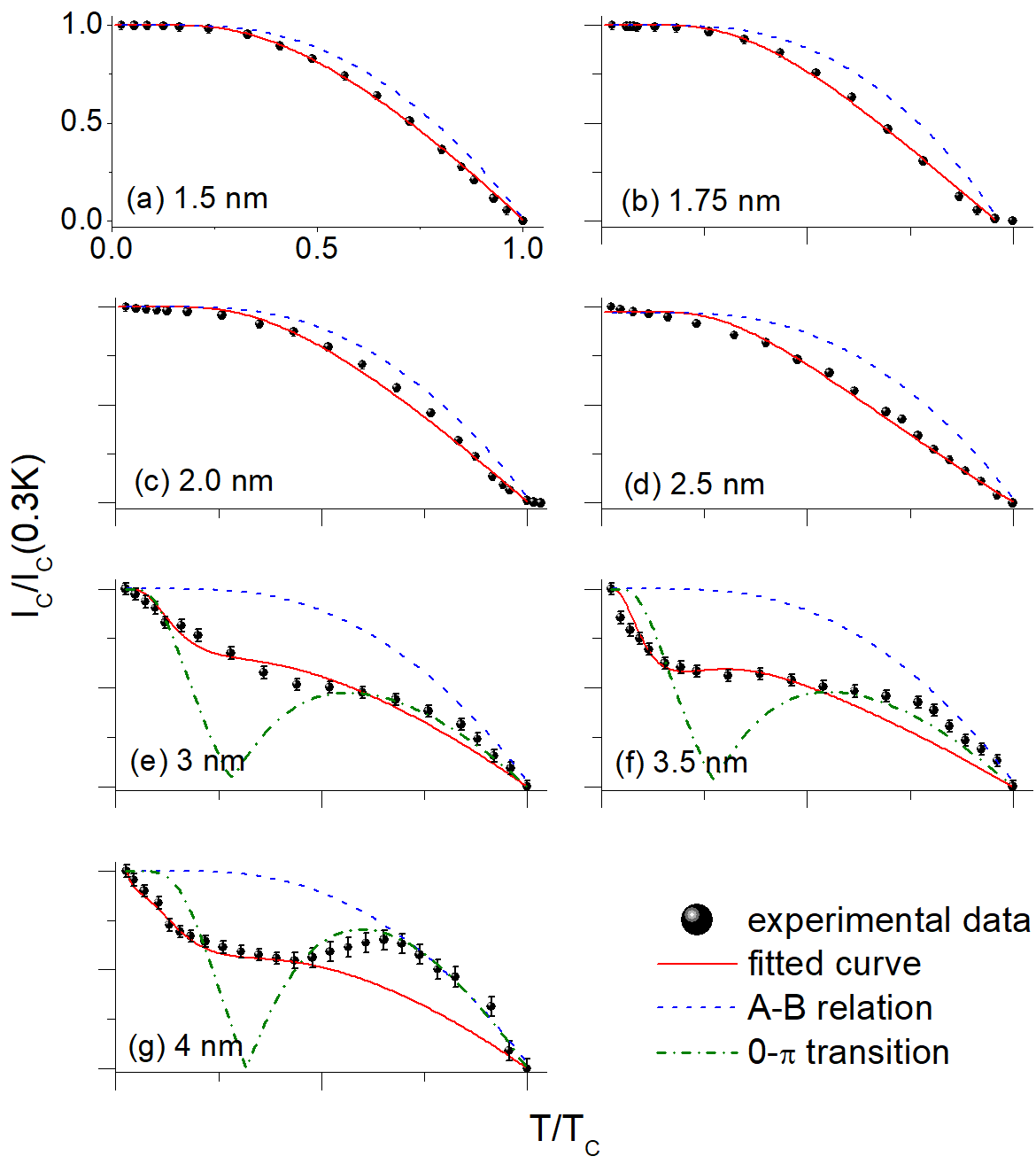}
\caption{Black dots: $I_C(T)$ for all samples. Blue dashed lines: Ambegaokar-Baratoff relation plotted for comparison. Red lines: fitted curves. In panel (a) to (d), Eq.\ref{eqn:cpr} is used. For panels from (e) to (g) Eq. \ref{eqn:ictsum} is used. Green lines: theoretical curves showing the $0-\pi$ transition obtained using Eq.\ref{eqn:cpr}. In all panels, $I_C$ is normalized to $I_C(0.3K)$, while $T$ is normalized to $T_C$ for each junction.}
\label{fig:fitIcSum}
\efig
The presence of a spin-active interface is supported by conductance spectra measurements (see Supplementary Material for details). Here we find a finite background conductance that increases as the barrier thickness increases, which is caused by an intrinsic asymmetry between the two interfaces NbN/GdN and GdN/NbN, due to the fabrication process. 
In $dI/dV$ measurements, the spin-filtering and the broadening due to finite Cooper pair life time has been modeled following Refs.\cite{Dynes:1978, Meservey:1970, Tedrow:1971, Tedrow:1986, Pal:2015}.

A more complete modeling of the data would require a detailed understanding of the micromagnetics of our barriers, including taking into account its multi-domain nature.
In the absence of this, we refrain from pushing our model too far, and only conjecture that a more realistic statistical treatement of the interface channels in combination of a self-consistent evaluation of the spin-triplet pair corrleations across the junction would account for the remaining differences between our model and the data.

The fitting parameters for short and long junctions are consistent with the underlying physics of these systems. The transparencies $t_{\uparrow}$ and $t_{\downarrow}$ decrease as the barrier thickness increases, as one would expect for tunnel junctions. The spin mixing angles for short junctions $1.5 $ nm $\leq d\leq2.5 $ nm are relatively small, indicating a moderate magnetic activity, and both the spin mixing angles used to fit longer junctions, with 3 nm$\leq d \leq$4 nm, are higher and close to $\pi$, confirming the stronger magnetic activity in these samples. Finally, the parameter $g$ is reasonable for the assumption of the presence of different transport channels with comparable weights inside the barrier. 

In conclusion, we have extensively characterized spin filter Josephson junctions in a wide range of temperatures.
Our measurements give clear indications on the occurrence of unconventional magnetic activity in spin filter Josephson junctions. They have also shown evidence of an incomplete $0-\pi$ transition in spin-filter Josephson junctions for the first time, further promoting the possible implementation of such junctions in a variety of applications, including those related to quantum circuits both as active \cite{davideNATCOM, Kawabata:2006} and passive $\pi-$shifter elements \cite{qbit:ustinov2}. 
This transition can only be described assuming non-uniform exchange interactions in the ferromagnetic barrier. This, combined with the observed large spin-filter efficiency, constitutes a strong indication of the presence of spin-triplet Cooper pairs strongly modifying the critical Josephson current.

R. Caruso, D. Massarotti and F. Tafuri thank NANOCOHYBRI project (Cost Action CA 16218), M. Eschrig and M. G. Blamire were supported by EPSRC Programme Grant EP/N017242/1.

\bibliography{nota}

\end{document}